\documentclass[12pt]{revtex4}
\usepackage{amsfonts}
\usepackage{graphicx, amsmath}
\usepackage{longtable}
\makeatletter

\newcommand{\Rmnum}[1]{\expandafter\@slowromancap\romannumeral #1@}
\makeatother
\begin{document}
\title[Short Title]{Fast adiabatic quantum state transfer and entanglement generation between two atoms via dressed states}
\author{Jin-Lei Wu}
\author{Xin Ji\footnote{E-mail: jixin@ybu.edu.cn}}
\author{Shou Zhang}
\affiliation{Department of Physics, College of Science, Yanbian University, Yanji, Jilin 133002, People's Republic of China}

\begin{abstract}Recently, a new method, which can significantly speed up adiabatic quantum state transfer by using dressed states, was proposed by Baksic \emph{et~al.} [Phys. Rev. Lett. \textbf{116}, 230503 (2016)]. Assisted by quantum Zeno dynamics, we develop this dressed-state method to achieve shortcuts to complete and fractional stimulated Raman adiabatic passage for speeding up adiabatic two-atom quantum state transfer and maximum entanglement generation, respectively. By means of some numerical simulations, we determine the parameters used in the scheme which can guarantee the feasibility and efficiency both in theory and experiment. Besides, we give strict numerical simulations to discuss the scheme's robustness, and the results show the scheme is robust against the variations in the parameters, atomic spontaneous emissions and the photon leakages from the cavity.
\pacs{42.50.Ct, 03.65.Ud, 32.80.Qk}
\end{abstract}

\maketitle
\section{Introduction}
Quantum state transfer and entanglement generation between different systems with time-dependent interacting fields have become more and more important for the further development of quantum information processing~\cite{KHB1998,JMM2006,H2008}. In order to achieve the high-fidelity quantum state transfer and entanglement generation, adiabatic evolution which corresponds to a state transfer along the adiabatic eigenstates is an excellent candidate method~\cite{NKB1999,NTB2001}. The most widely used approach of adiabatic evolution is stimulated Raman adiabatic passage~(STIRAP) because of its great robustness against pulse area and timing errors as well as the restraint on lossy intermediate states.

However, STIRAP schemes usually require a relatively long interaction time, and thus the adiabatic evolution may suffer from dissipation and noise during the process of quantum state transfer. Therefore, it is greatly necessary to speed up the process of adiabatic evolution and lots of techniques have been brought forward~\cite{XIA2010,SSX2011,SXE2012,ESS2013,A2013,YYQ2016,XQJ2016}. There are two techniques, transitionless quantum driving and Lewis-Riesenfeld invariants, widely applied to speed up adiabatic quantum state transfer and entanglement generation~\cite{MS2003,MS2008,M2009,HW1969,XJ2012,LE2014,SS2015}. Although, in theory, the high-fidelity adiabatic quantum state transfer and entanglement generation can be achieved by transitionless quantum driving and Lewis-Riesenfeld invariants, it is hardly possible in practice due to the major flaws of the two techniques. On one hand, a transitionless-based direct coupling between the initial state and the target state is too hard to be obtained experimentally~\cite{MYL2014,YYJ2015,ZYY2016}. On the other hand, invariants-based driving pulses are usually not smoothly turned on or off and thus lead to severe impediments in experiment~\cite{YYQ2014,YYQ2015,YLQ2015,YLC2015,YLX2015}.

A short time before, Baksic \emph{et~al.} proposed a new method to speed up adiabatic quantum state transfer by using dressed states~\cite{AHA2016}. In Ref.~\cite{AHA2016}, the dressed states are skillfully defined to incorporate the nonadiabatic processes. Inspired by Ref.~\cite{AHA2016}, we apply the dressed-state method to quantum state transfer and entanglement generation between two $\Lambda$-type atoms trapped in an optical cavity. With the assist of quantum Zeno dynamics~\cite{PS2002,PGS2009}, the original system's Hamiltonian is greatly simplified and viewed as a three-level system. With the addition of a suitable correction Hamiltonian to the original Hamiltonian and the ingenious unitary transformation, we construct a modified diagonal adiabatic Hamiltonian possessing a set of dressed states as its adiabatic eigenstates. By finding out the appropriate parameters, we implement perfectly the fast adiabatic two-atom quantum state transfer and entanglement generation with the evolutions along one of the dressed states.

This paper is organized as follows. In Sec.~\ref{a}, we describe the two-atom system and show the dressed-state method to implement adiabatic two-atom quantum state transfer and entanglement generation. In Sec.~\ref{b}, we will give the numerical simulations for selecting the related parameters and discussing the scheme's effectiveness and robustness. The conclusion appears in Sec.~\ref{c}.

\section{Adiabatic quantum state transfer and entanglement generation between two atoms}\label{a}
\begin{figure}[ht]
\includegraphics[width=\linewidth]{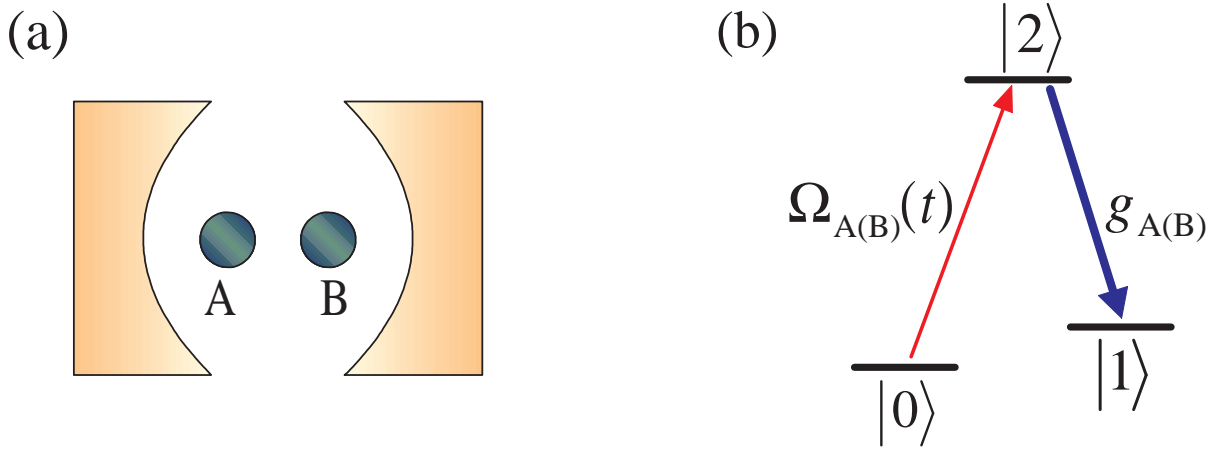}\\
\caption{(Color online) (a) The diagrammatic sketch of cavity-atom combined system. (b) Atomic level configuration.}\label{F1}
\end{figure}
The schematic setup for quantum state transfer and entanglement generation between two atoms is shown in Fig.~\ref{F1}. Two $\Lambda$-type atoms A and B are trapped in a single-mode optical cavity. Each atom has an excited state $|2\rangle$ and two ground states $|0\rangle$ and $|1\rangle$. The atomic transition $|2\rangle_{\rm A(B)}\leftrightarrow|1\rangle_{\rm A(B)}$ is resonantly coupled to the mode of the cavity with corresponding coupling constant $g_{\rm A(B)}$, and the transition $|2\rangle_{\rm A(B)}\leftrightarrow|0\rangle_{\rm A(B)}$ is resonantly driven by classical field with the time-dependent Rabi frequency $\Omega_{\rm A(B)}(t)$. Then the atom-cavity system can be dominated by the interaction Hamiltonian~(setting $\hbar=1$)
\begin{eqnarray}\label{e1}
H_{\rm total}(t)&=&H_{al}(t)+H_{ac}(t),\nonumber\\
H_{al}(t)&=&\sum_{k=\rm{A, B}}\Omega_{k}(t)|2\rangle_{k}\langle0|+\mathrm{H.c.},\nonumber\\
H_{ac}(t)&=&\sum_{k=\rm{A, B}}g_{k}a|2\rangle_{k}\langle1|+\mathrm{H.c.},
\end{eqnarray}
where $H_{al}(t)~(H_{ac}(t))$ is the interaction between the atoms and the classical laser fields~(the mode of the cavity), and $a$ is the annihilation operator of the cavity mode. For simplicity, we assume $g_{\rm A}=g_{\rm B}=g$. Then suppose that the total system is initially in the state $|\phi_1\rangle=|0\rangle_{\rm A}|1\rangle_{\rm B}|0\rangle_{\rm c}$ denoting atom A, atom B and the cavity mode in the state $|0\rangle_{\rm A}$, state $|1\rangle_{\rm B}$ and vacuum state, respectively. Thus dominated by the Hamiltonian~(\ref{e1}), the whole system evolves in the Hilbert space spanned by
\begin{eqnarray}\label{e2}
|\phi_1\rangle&=&|0\rangle_{\rm A}|1\rangle_{\rm B}|0\rangle_{\rm c},\nonumber\\
|\phi_2\rangle&=&|2\rangle_{\rm A}|1\rangle_{\rm B}|0\rangle_{\rm c},\nonumber\\
|\phi_3\rangle&=&|1\rangle_{\rm A}|1\rangle_{\rm B}|1\rangle_{\rm c},\nonumber\\
|\phi_4\rangle&=&|1\rangle_{\rm A}|2\rangle_{\rm B}|0\rangle_{\rm c},\nonumber\\
|\phi_5\rangle&=&|1\rangle_{\rm A}|0\rangle_{\rm B}|0\rangle_{\rm c}.
\end{eqnarray}
Obviously, the system is initially in the dark state of $H_{ac}(t)$, i.e., $H_{ac}(t)|\phi_1\rangle=0$. Then choosing the quantum Zeno limit condition $\Omega_{\rm A}(t), \Omega_{\rm B}(t)\ll g$, the whole system can approximatively evolve in an invariant Zeno subspace consisting of dark states of $H_{ac}(t)$~\cite{XHL2010,XLS2009}
\begin{eqnarray}\label{e3}
H_P=\Big\{|\phi_1\rangle, |\phi_d\rangle, |\phi_5\rangle \Big\},
\end{eqnarray}
corresponding to the projections
\begin{equation}\label{e4}
P^\alpha=|\alpha\rangle\langle\alpha|,\quad(|\alpha\rangle\in H_{P}).
\end{equation}
Here,
\begin{eqnarray}\label{e5}
|\phi_d\rangle&=&\frac{1}{\sqrt{2}}\Big(-|\phi_2\rangle+|\phi_4\rangle\Big).
\end{eqnarray}
Therefore, the system Hamiltonian can be rewritten as the following form~\cite{XLS2010}
\begin{eqnarray}\label{e6}
H(t)&\simeq&\sum_{\alpha}P^\alpha H_{al}(t)P^\alpha~\nonumber\\
&=&\Omega_1(t)|\phi_1\rangle\langle\phi_d|+\Omega_2(t)|\phi_5\rangle\langle\phi_d|+\rm H.c.,
\end{eqnarray}
in which $\Omega_1(t)=-\Omega_{\rm A}(t)/\sqrt{2}$ and $\Omega_2(t)=\Omega_{\rm B}(t)/\sqrt{2}$. There are three time-dependent eigenstates of the Hamiltonian~(\ref{e6}), $|\varphi_d(t)\rangle$ and $|\varphi_{\pm}(t)\rangle$~(see the APPENDIX), corresponding to the eigenvalues $E_d=0$ and $E_{\pm}=\pm \Omega(t)$ ($\Omega(t)=\sqrt{\Omega_1(t)^2+\Omega_2(t)^2}$), respectively.

It is convenient to move the time-dependent adiabatic frame to the time-independent adiabatic frame by the unitary operator $U(t)=\sum_{j=d,\pm}|\varphi_j\rangle\langle\varphi_j(t)|$. In the time-independent adiabatic frame, there are still nonadiabatic couplings in the Hamiltonian, which may lead to an imperfect state transfer~(see the APPENDIX). In order to correct the nonadiabatic errors, we look for a correction Hamiltonian $H_c(t)$ such that the modified Hamiltonian $H_{\rm mod}(t)=H(t)+H_c(t)$ governs a perfect state transfer. Here we choose the general form of $H_c(t)$
\begin{eqnarray}\label{e7}
H_{c}(t)=U^\dag(t)(g_x(t)M_x+g_z(t)M_z)U(t),
\end{eqnarray}
where $M_x=(|\varphi_-\rangle-|\varphi_+\rangle)\langle\varphi_d|/\sqrt{2}+\rm H.c.$ and $M_z=|\varphi_+\rangle\langle\varphi_+|-|\varphi_-\rangle\langle\varphi_-|$. Thus we obtain the modified pulses
\begin{eqnarray}\label{e8}
\Omega'_1(t)=g_x(t)\cos\theta(t)-[g_z(t)+\Omega(t)]\sin\theta(t),\nonumber\\
\Omega'_2(t)=g_x(t)\sin\theta(t)+[g_z(t)+\Omega(t)]\cos\theta(t),
\end{eqnarray}
with $\theta(t)=\arctan(\Omega_1(t)/\Omega_2(t))$.

Now, we define a new basis of dressed states $|\tilde{\varphi}_{\pm,d}(t)\rangle$ by the action of a time-dependent unitary operator $V(t)$ on the time-independent eigenstates $|\varphi_{\pm,d}\rangle$, i.e., $|\tilde{\varphi}_{\pm,d}(t)\rangle=V(t)|\varphi_{\pm,d}\rangle$. In this scheme, we choose
\begin{eqnarray}\label{e9}
V(t)=\exp[i\mu(t)M_x],
\end{eqnarray}
with an Euler angle $\mu(t)$. Then moving the modified Hamiltonian $H_{\rm mod}$ to the frame defined by $V(t)$ and choosing the appropriate control parameters, we can obtain a new Hamiltonian $H_{\rm new}(t)=V H_{\rm ad}(t)V^\dag+VU H_c(t)U^\dag V^\dag+i\dot{V}V^\dag$ which is with no unwanted off-diagonal elements~(see the APPENDIX).

Back to the time-dependent adiabatic frame, the dark dressed state of $H_{\rm new}(t)$, the eigenstate with the eigenvalue $\tilde{E}_{d}=0$, is written as
\begin{eqnarray}\label{e10}
|\tilde{\varphi}_{d}(t)\rangle&=&\cos\mu(t)\Big[\cos\theta(t)|\phi_1\rangle+\sin\theta(t)|\phi_5\rangle\Big]\nonumber\\
&&+i\sin\mu(t)|\phi_d\rangle.
\end{eqnarray}
Obviously, the dark dressed state $|\tilde{\varphi}_{d}(t)\rangle$ can serve as a medium state which operates a two-atom quantum state transfer $|\phi_1\rangle\rightarrow|\phi_5\rangle$ by setting $\theta(t_i)=0$, $\theta(t_f)=\pi/2$ and $\mu(t_i)=\mu(t_f)=0$, where $t_{i(f)}$ is the initial~(final) time. Analogously, based on the fractional STIRAP proposed by Vitanov \emph{et~al.}~\cite{NKB1999}, a maximum two-atom entangled state $|\Psi\rangle=\frac{1}{\sqrt{2}}(|\phi_1\rangle+|\phi_5\rangle)=\frac{1}{\sqrt{2}}(|0\rangle_{\rm A}|1\rangle_{\rm B}+|1\rangle_{\rm A}|0\rangle_{\rm B})$ can be generated by setting $\theta(t_i)=0$, $\theta(t_f)=\pi/4$ and $\mu(t_i)=\mu(t_f)=0$.

Based on the process above, we have developed the dressed-state method to achieve shortcuts to complete and fractional STIRAP for speeding up adiabatic two-atom quantum state transfer and maximum entanglement generation, respectively. The evolution process is not necessarily slow and there is no direct coupling between the initial state and the target state, as long as a set of suitable dressed states is chosen.

\section{Numerical simulations}\label{b}
\subsection{Selections of parameters}
\begin{figure}[ht]
\begin{center}
\includegraphics[width=\linewidth]{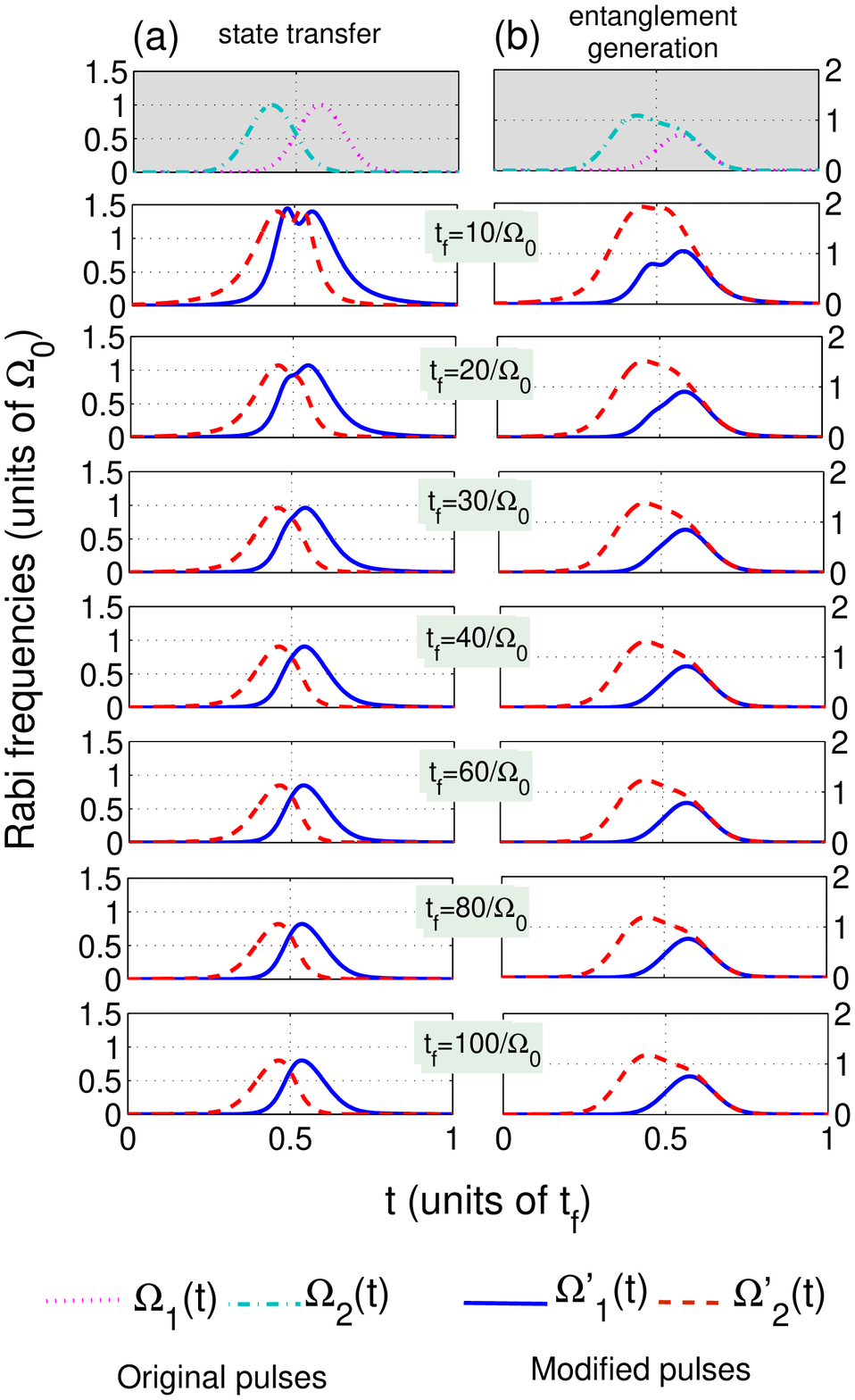}\\
\caption{(Color online) The shapes of the modified pulses $\Omega'_1(t)$ and $\Omega'_2(t)$ with several different values of the operation time $t_f$. The parameters used here are $t_0=3t_f/40$, $\tau=0.1t_f$, $\theta(t_f)=\pi/2$ for the quantum state transfer or $\theta(t_f)=\pi/4$ for the entanglement generation.}\label{F2}
\end{center}
\end{figure}

\begin{figure}[htb]
\begin{center}
\includegraphics[width=\linewidth]{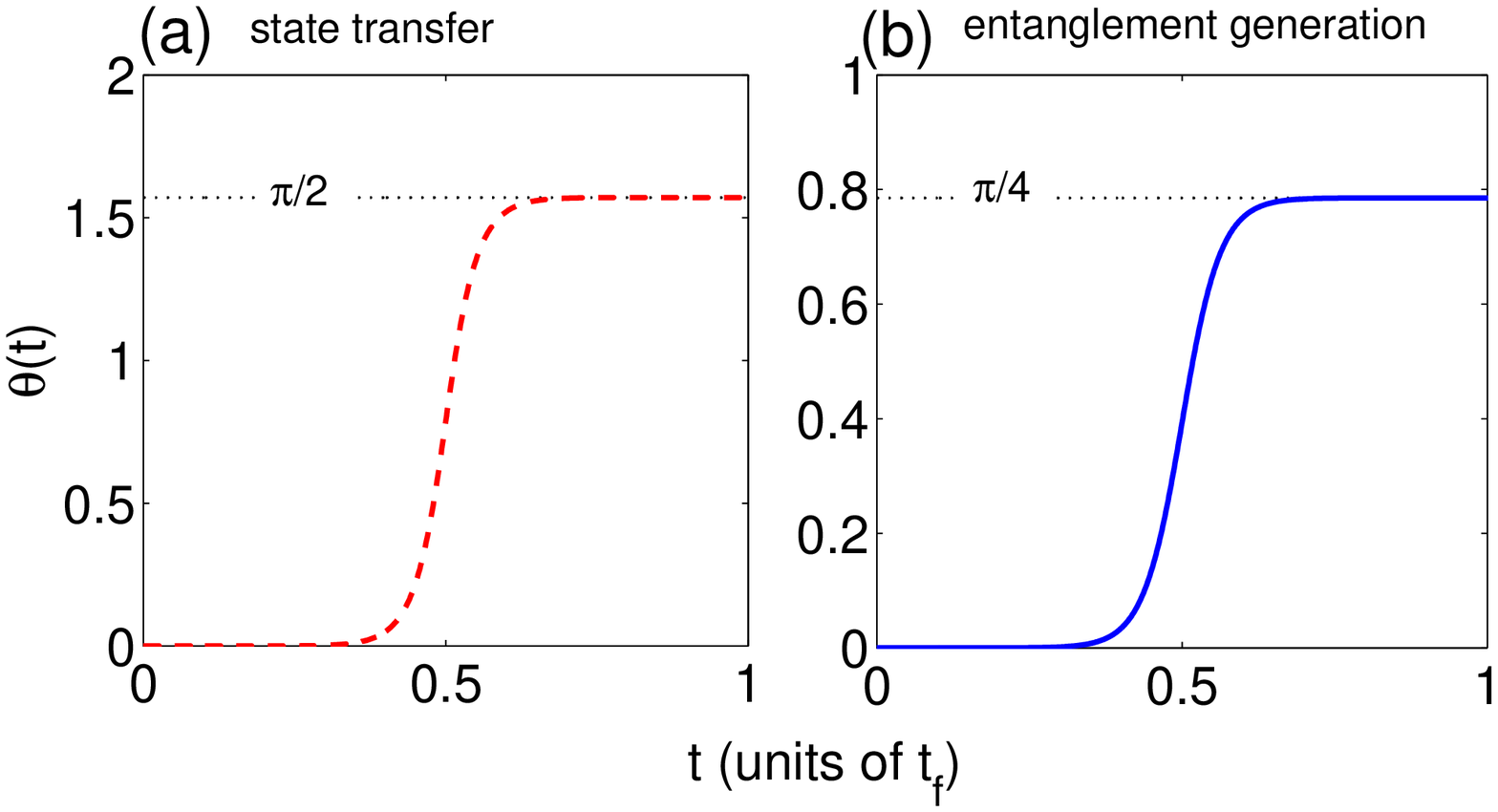}\\
\caption{(Color online) Time dependence of $\theta(t)$ with an arbitrary $t_f$. The parameters used here are the same as in Fig.~\ref{F2}.}\label{F3}
\end{center}
\end{figure}

First of all, we give the numerical simulations to select appropriate parameters for insuring the experimental and theoretical feasibility. The original pulses $\Omega_1(t)$ and $\Omega_2(t)$ can be chosen as the Gaussian pulses~\cite{KHB1998,NKB1999}
\begin{eqnarray}\label{e11}
\Omega_1(t)&=&\sin\theta(t_f)\Omega_0\exp[-(t-t_f/2-t_0)^2/\tau^2],\nonumber\\
\Omega_2(t)&=&\cos\theta(t_f)\Omega_0\exp[-(t-t_f/2-t_0)^2/\tau^2]\nonumber\\
&&+\Omega_0\exp[-(t-t_f/2+t_0/2)^2/\tau^2],
\end{eqnarray}
and the corresponding $\Omega_{\rm A}(t)=-\sqrt{2}\Omega_1(t)$ and $\Omega_{\rm B}(t)=\sqrt{2}\Omega_2(t)$, the original driving pulses of the total system, can be obtained. In the scheme, we choose the pulses' time delay $t_0=3t_f/40$ and width $\tau=0.1t_f$ as functions with respect to the operation time $t_f$ (the initial time $t_i=0$), respectively. Besides, $\theta(t_f)=\pi/2$ for the quantum state transfer and $\theta(t_f)=\pi/4$ for the entanglement generation, respectively. The Euler angle $\mu(t)$ is defined by~\cite{AHA2016}
\begin{eqnarray}\label{e12}
\mu(t)=-\arctan\Big(\frac{\dot{\theta}(t)}{G(t)/\tau+\Omega(t)}\Big),
\end{eqnarray}
where $G(t)=\mathrm{sech}(t/\tau)$ is chosen to regularize $\mu(t)$ such that it can meet the condition $\mu(t_i)=\mu(t_f)=0$ and make $\sin^2\mu(t)$, the population of $|\phi_d\rangle$~(see Eq.~(\ref{e10})), as small as possible. Then based on the relevant parameters above, the modified pulses $\Omega'_1(t)$ and $\Omega'_2(t)$ can be determined by Eq.~(\ref{e8}). Also, the corresponding modified driving pulses of the total system $\Omega'_{\rm A}(t)=-\sqrt{2}\Omega'_1(t)$ and $\Omega'_{\rm B}(t)=\sqrt{2}\Omega'_2(t)$ are obtained.

\begin{figure}[htb]
\begin{center}
\includegraphics[width=\linewidth]{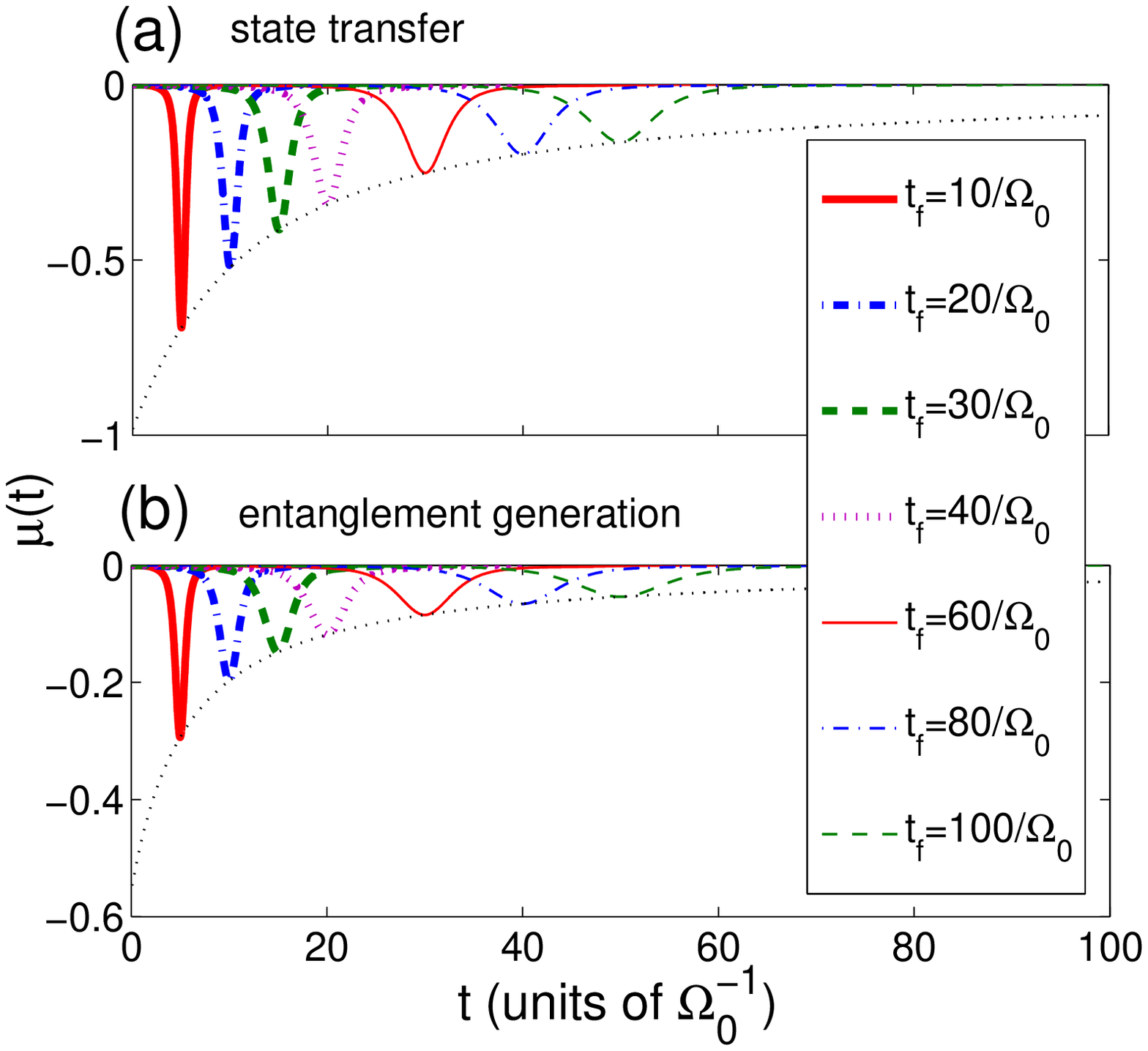}\\
\caption{(Color online) Time dependence of $\mu(t)$ with several different values of the operation time $t_f$. The parameters used here are the same as in Fig.~\ref{F2}.}\label{F4}
\end{center}
\end{figure}
In order to choose a small and feasible value of the operation time $t_f$, in Fig.~\ref{F2}, we plot the shapes of the modified pulses $\Omega'_1(t)$ and $\Omega'_2(t)$ with different values of $t_f$. From Fig.~\ref{F2}, we can clearly find that, both for the quantum state transfer and the entanglement generation, the longer the operation time $t_f$ is, the more similar the shapes of the modified pulses are to those of the original pulses. In other words, the modified pulses' experimental feasibility increases as the operation time $t_f$ increases, which implies the operation time we choose can not be too short. Then, taking into account the boundary conditions $\theta(t_i)=0$ and $\theta(t_f)=\pi/2$ for the quantum state transfer or $\theta(t_f)=\pi/4$ for the entanglement generation, we plot the time dependence of $\theta(t)$ in Fig.~\ref{F3} with an arbitrary $t_f$. Without a doubt, Fig.~\ref{F3} shows that the boundary conditions with respect to $\theta$ can be satisfied perfectly, and the time dependence of $\theta(t)$ are independent of $t_f$. For the boundary condition $\mu(t_i)=\mu(t_f)=0$, we show it in Fig.~\ref{F4} by plotting the time dependence of $\mu(t)$ with several different values of $t_f$. Apparently, the boundary condition $\mu(t_i)=\mu(t_f)=0$ is always satisfied well with an arbitrary $t_f$. The maximum values of $|\mu(t)|$, however, always decrease with the increase of $t_f$. Therefore, for $|\phi_d\rangle$'s population $\sin^2\mu(t)$ to be small enough, the operation time we choose can not be too short.

\begin{figure}[htb]
\begin{center}
\includegraphics[width=\linewidth]{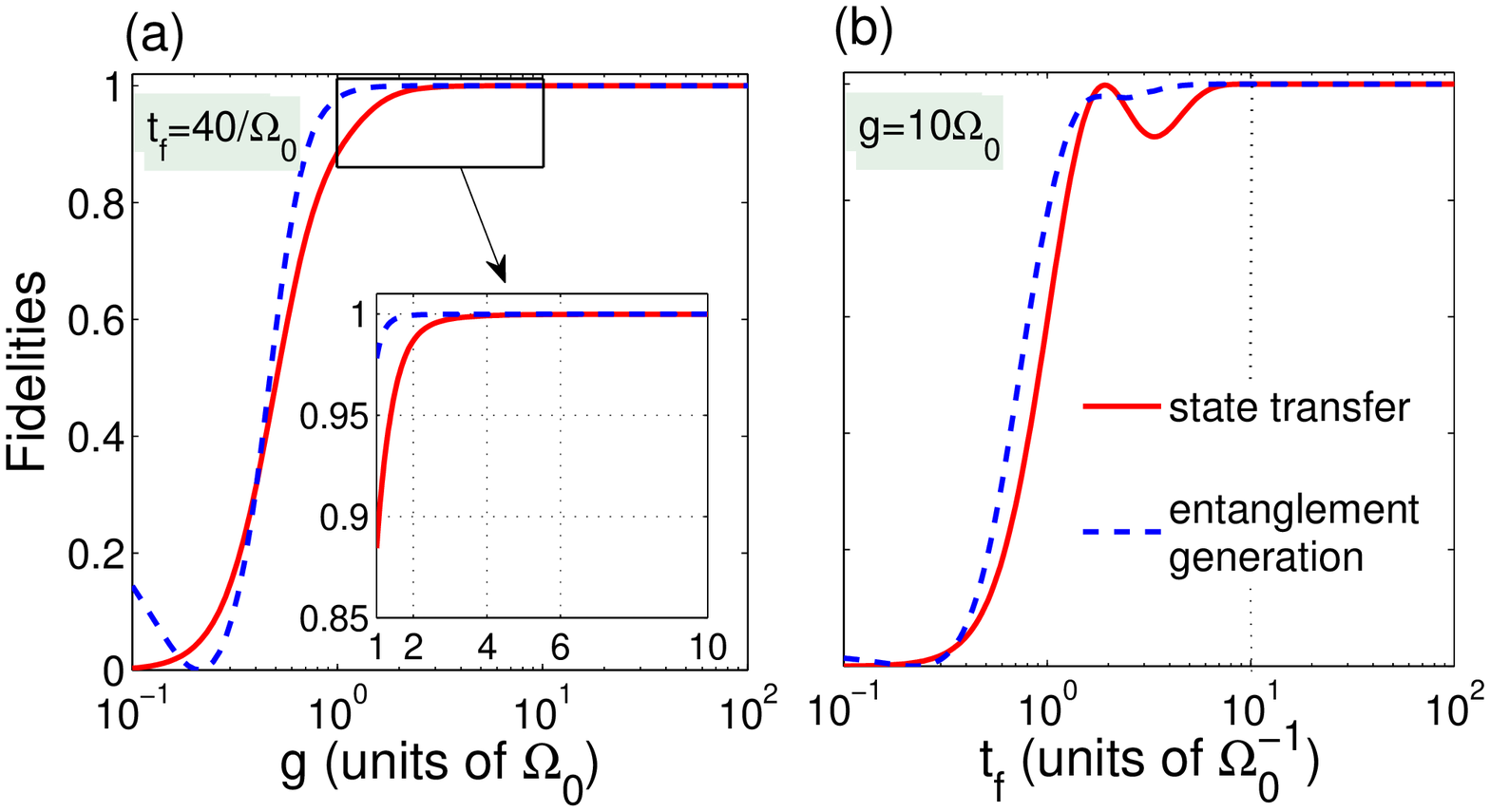}\\
\caption{(Color online) (a)~Fidelities as functions of $g$ with $t_f=40/\Omega_0$; (b)~fidelities as functions of $t_f$ with $g=10\Omega_0$. Other parameters used here are the same as in Fig.~\ref{F2}.}\label{F5}
\end{center}
\end{figure}

For the high experimental feasibility of the scheme and a relatively small occupancy of $|\phi_d\rangle$, we preselect $t_f=40/\Omega_0$. Then with $t_f=40/\Omega_0$, in Fig.~\ref{F5}(a), we plot the fidelities of the dressed-state scheme as functions of the atom-cavity coupling strength $g$, where the fidelities are defined by $F=|\langle\phi_{ideal}|\phi(t)\rangle|^2$ with $|\phi_{ideal}\rangle=|\phi_5\rangle$ for the state transfer or $|\phi_{ideal}\rangle=\frac{1}{\sqrt{2}}(|\phi_1\rangle+|\phi_5\rangle)$ for the entanglement generation, respectively. $|\phi(t)\rangle$ is the state of system governed by the modified Hamiltonian. Because of the Zeno limit condition $\Omega_{\rm A}(t), \Omega_{\rm B}(t)\ll g$, the fidelities increase with the increase of $g/\Omega_0$. However, as shown in Fig.~\ref{F5}(a), even when $g=4\Omega_0$ which does not strictly meet the Zeno limit condition, the fidelities are almost unit both for the quantum state transfer and the entanglement generation. Here we preselect $g=10\Omega_0$ to guarantee the scheme's robustness. For checking the feasibility of the preselection $t_f=40/\Omega_0$, we plot the fidelities as functions with respect to $t_f$ in Fig.~\ref{F5}(b) with $g=10\Omega_0$, and Fig.~\ref{F5}(b) indicates that the preselection $t_f=40/\Omega_0$ is feasible and robust both for the quantum state transfer and the entanglement generation.
\subsection{Discussion of effectiveness}
\begin{figure}[htb]
\begin{center}
\includegraphics[width=\linewidth]{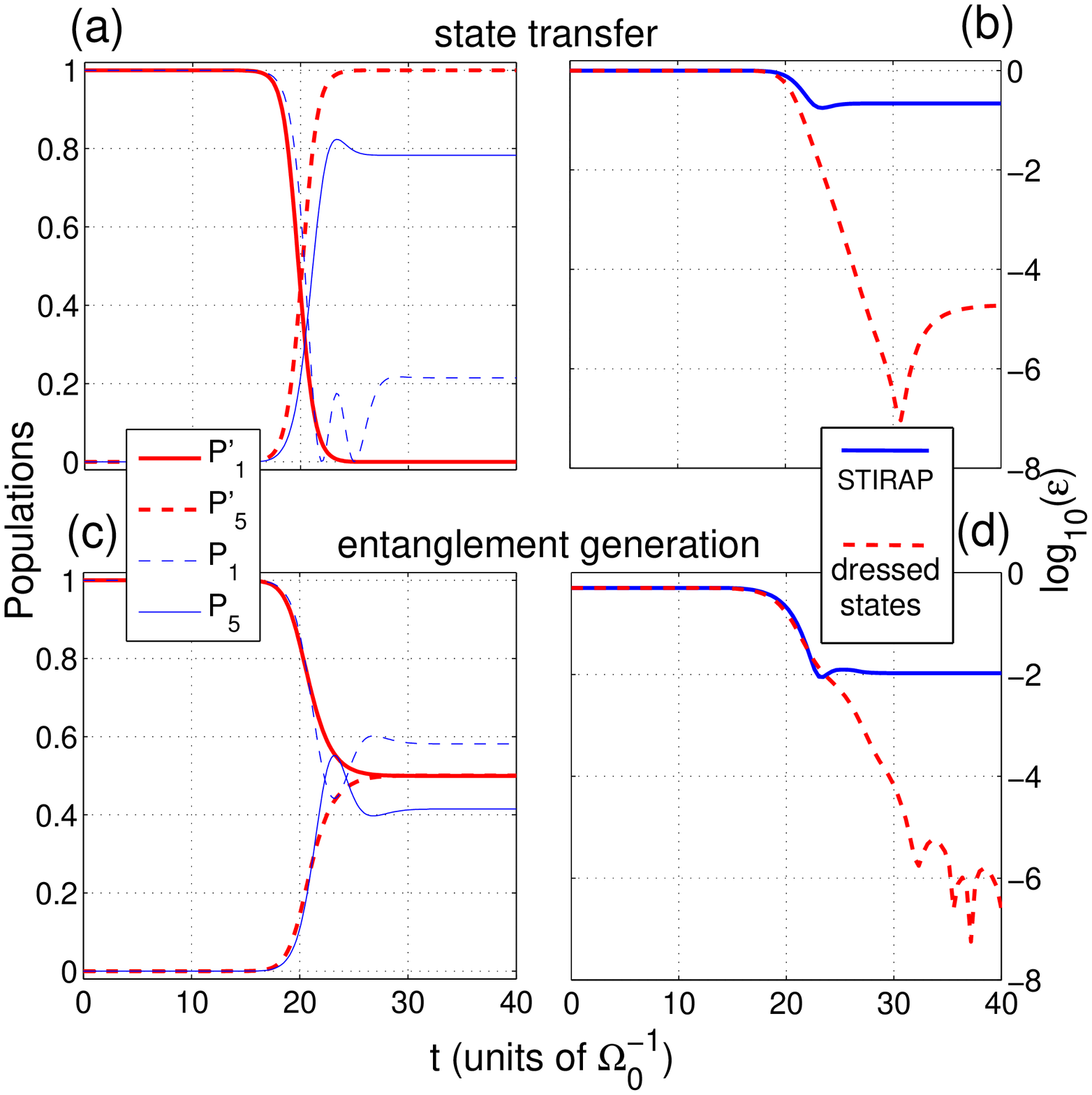}\\
\caption{(Color online) (a) and (c): time dependence of the populations of the states $|\phi_1\rangle$ and $|\phi_5\rangle$ based on the dressed-state scheme~(thick red lines) and the STIRAP scheme~(thin blue lines); (b) and (d): residual errors of the states based on the dressed-state scheme~(red dashed lines) and the STIRAP scheme~(blue solid lines). $g=10\Omega_0$, $t_f=40/\Omega_0$ and other parameters used here are the same as in Fig.~\ref{F2}.}\label{F6}
\end{center}
\end{figure}

Next, in order to show the dressed-state scheme's effectiveness, in Fig.~\ref{F6}, we show the time dependence of the populations of the states $|\phi_1\rangle$ and $|\phi_5\rangle$ and the residual errors $\varepsilon(t)=1-|\langle\phi_{ideal}|\phi(t)\rangle|^2$ of the quantum state transfer and the entanglement generation,respectively. Here, $|\phi(t)\rangle$ is the state of the total system based on the dressed-state scheme~(with the driving pulses $\Omega'_{\rm A}(t)$ and $\Omega'_{\rm B}(t)$) or the STIRAP scheme~(with the driving pulses $\Omega_{\rm A}(t)$ and $\Omega_{\rm B}(t)$). As shown in Figs.~\ref{F6}(a) and \ref{F6}(c), the dressed-state scheme achieves the perfect desired population transfer both for the quantum state transfer and the entanglement generation, but the STIRAP scheme can not perfectly achieve the quantum state transfer or the entanglement generation. Correspondingly, in Figs.~\ref{F6}(b) and \ref{F6}(d), the dressed-state scheme leads to a reduction of the residual errors by over four orders of magnitude at the final time both for the state transfer and the entanglement generation. Therefore, there is no doubt that the dressed-state scheme is highly feasible and effective even within a very short operation time.

\subsection{Discussion of robustness}
In the above discussion, we think the operations and the whole system perfect and absolutely isolated from the environment. Therefore, it is necessary to give the discussions about the robustness of the scheme against the pulse parameters' imperfections, as well as the variations in the parameters and decoherence induced by the atomic spontaneous emissions and photon leakages from the cavity.

\begin{figure}[htb]
\begin{center}
\includegraphics[width=\linewidth]{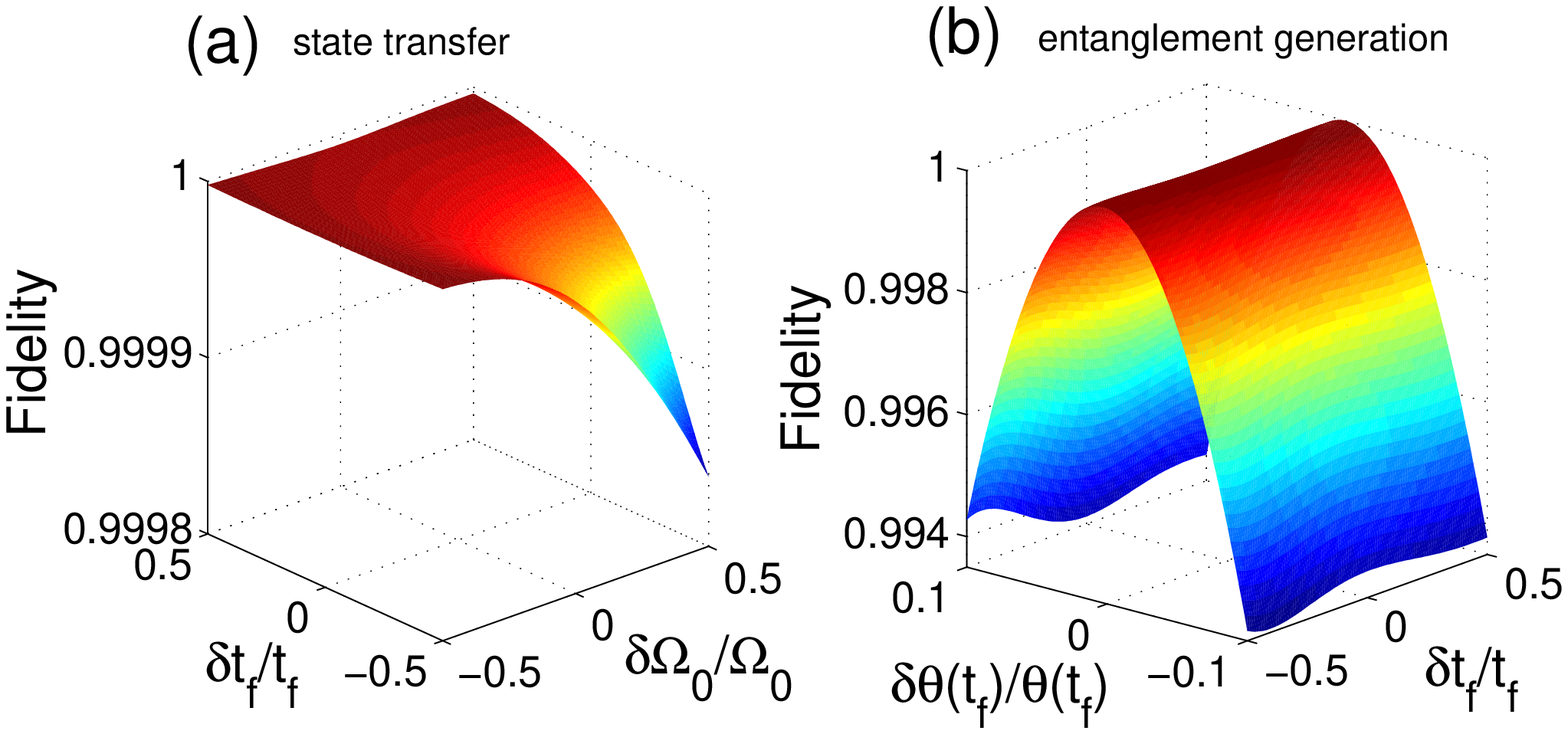}\\
\caption{(Color online) (a)~The fidelity versus the variations in the pulses time $t_f$ and amplitude $\Omega_0$ for the state transfer; (b)~the fidelity versus the variations in $t_f$ and $\theta(t_f)$ for the entanglement generation. The parameters used here are same as in Fig.~\ref{F6}.}\label{F7}
\end{center}
\end{figure}
We first consider the robustness of the pulse sequences by plotting the fidelity versus the variations in the pulses time $t_f$ and amplitude $\Omega_0$ in Fig.~\ref{F7}(a) for the two-atom state transfer based the dressed-state scheme. Here we define $\delta x= x'-x$ as the deviation of $x$, in which $x$ denotes the ideal value and $x'$ denotes the actual value. From Fig.~\ref{F7}(a), we learn that the fidelity slightly decreases with the increasing amplitude $\Omega_0$. It is clear that the increase of $\Omega_0$ causes the increase of the modified pulses' amplitudes, and thus, to a certain extent, the Zeno limit condition will be spoiled. While simultaneously we can also see that, the longer $t_f$ is, the higher the fidelity is. The reason can be deduced from Fig.~\ref{F2} that when the operation time increases, the amplitudes of the modified pulses will decrease, and thus the Zeno limit condition will be met more strictly. Taking one with another, however, the fidelity always keeps near $F=1$, and hence the dressed-state scheme have the extremely high robustness against the variations in $t_f$ and $\Omega_0$.
Analogously, the entanglement generation has a similar situation. Dissimilarly, however, the condition $\sin\theta(t_f)=\cos\theta(t_f)$~(i.e., $\theta(t_f)=\pi/4$) in Eq.~(\ref{e11}) is very critical for the maximum entanglement generation. Therefore, we are supposed to consider the effect of the variations in $\theta(t_f)$ on the fidelity. We plot the fidelity versus the variations in $t_f$ and $\theta(t_f)$ in Fig.~\ref{F7}(b) for the entanglement generation based on the dressed-state scheme. To all appearances, in Fig.~\ref{F7}(b), the effect of the variations in $\theta(t_f)$ on the fidelity is far greater than that of the variations in $t_f$. But even so, the fidelity can keep quite high even when $|\delta\theta(t_f)/\theta(t_f)|=0.1$. To sum up, the dressed-state scheme are robust against variations in the parameters of the pulse sequences.

\begin{figure}[htb]
\begin{center}
\includegraphics[width=\linewidth]{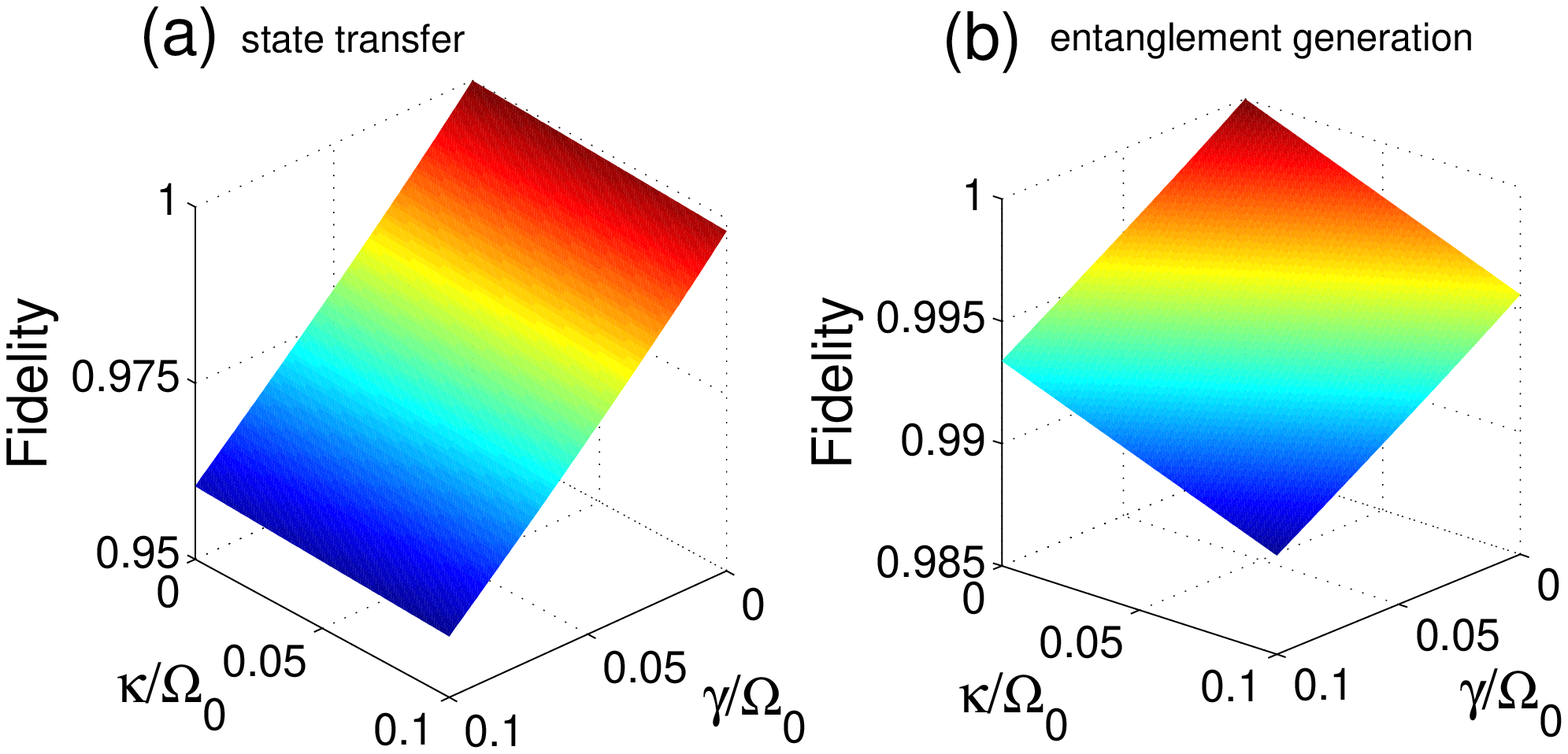}\\
\caption{(Color online) The fidelity as a function of $\gamma/\Omega_{0}$ and $\kappa/\Omega_{0}$. The parameters used here are the same as in Fig.~\ref{F6}.}\label{F8}
\end{center}
\end{figure}
Next, we take the decoherence induced by the atomic spontaneous emissions and the photon leakage from the cavity into account. Then the whole system is dominated by the master equation
\begin{eqnarray}\label{e13}
\dot{\rho}(t)&=&-i[H^{\rm mod}_{\rm total}(t),\rho(t)]\nonumber\\
\cr&&-\sum_{j=\rm A,B}\sum_{i=0,1}\frac{\gamma_{i}^{j}}{2}\Big(\sigma_{2,2}^{j}\rho-2\sigma_{i,2}^{j}\rho\sigma_{2,i}^{j}+\rho\sigma_{2,2}^{j}\Big)\nonumber\\
\cr&&-\frac{\kappa}{2}\Big(a^{\dag}a\rho-2a\rho a^{\dag}+\rho a^{\dag}a\Big),
\end{eqnarray}
where $H^{\rm mod}_{\rm total}(t)=\sum_{j=\rm{A, B}}(\Omega'_{j}(t)|2\rangle_{j}\langle0|+g_{j}a|2\rangle_{j}\langle1|)+\mathrm{H.c.}$; $\gamma_{i}^{j}$ is the spontaneous emission rate of $j$th atom from the excited state $|2\rangle_j$ to the ground state $|i\rangle_j$; $\kappa$ denotes the photon leakage rate from the cavity; $\sigma_{mn}^{j}=|m\rangle_{j}\langle n|$. For simplicity, we assume $\gamma_{i}^{j}=\gamma$. By means of the master equation, we plot the fidelities for the state transfer and the entanglement generation versus $\kappa/\Omega_0$ and $\gamma/\Omega_0$ in Fig.~\ref{F8}. Firstly, in Fig.~\ref{F8}, we can see that the fidelities for the state transfer and the entanglement generation are over 0.95 and 0.985, respectively, even when $\kappa=\gamma=0.1\Omega_0$. Without a doubt, the dressed-state scheme for the state transfer or the entanglement generation is robust against the decoherence induced by the atomic spontaneous emissions and the photon leakage from the cavity. In addition, as seen from the decrease of the fidelities with the increases of $\kappa$ and $\gamma$ in Fig.~\ref{F8}, we learn that the influence of the atomic spontaneous emissions on the fidelity is obviously greater than that of the photon leakages from the cavity. Even in Fig.~\ref{F8}(a), for the state transfer, the influence of the atomic spontaneous emissions on the fidelity almost plays a full role, but that of the photon leakages from the cavity is little. It follows that the dressed-state scheme we propose has the robustness against the photon leakages from the cavity more than the atomic spontaneous emissions.

\begin{figure}[htb]
\begin{center}
\includegraphics[width=\linewidth]{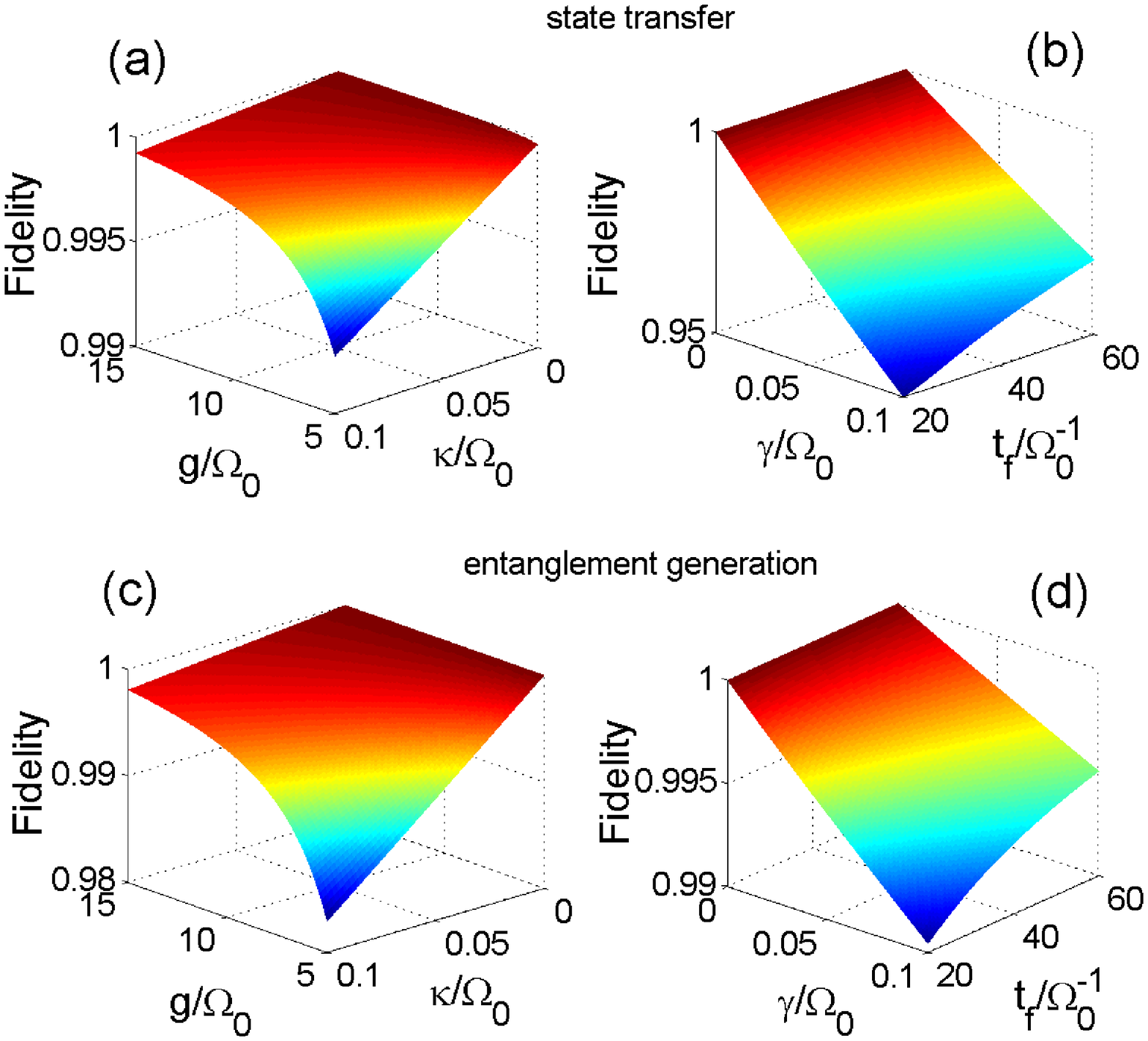}\\
\caption{(Color online) (a) and (c): the fidelities as functions of $g/\Omega_{0}$ and $\kappa/\Omega_{0}$ with $t_f=40/\Omega_0$ and $\gamma=0$; (b) and (d): the fidelities as functions of $t_f/\Omega^{-1}_{0}$ and $\gamma/\Omega_{0}$ with $g=10\Omega_0$ and $\kappa=0$. Other parameters used here are the same as in Fig.~\ref{F2}.}\label{F9}
\end{center}
\end{figure}

We have chosen the Zeno limit condition $\Omega_{\rm A}(t), \Omega_{\rm B}(t)\ll g$ to restrain the population of the cavity-mode excited state $|\phi_3\rangle$. Besides, we have known that the increase of the operation time $t_f$ leads to the decrease of $|\mu(t)|_{\rm max}$ from Fig.~\ref{F4}, and thus leads to the decrease of $|\phi_d\rangle$'s population $\sin^2\mu(t)$. Therefore, the influence of the photon leakages from the cavity and the atomic spontaneous emissions on the fidelity should be restrained by a bigger $g$ and a longer $t_f$, respectively. Based on this, we plot the fidelities as the functions with respect to $g$ and $\kappa$ in Figs.~\ref{F9}(a) and \ref{F8}(c) for the state transfer and the entanglement generation, respectively. Clearly, when the photon leakages from the cavity exist, a bigger value of $g$ can greatly depress the influence of the photon leakages from the cavity on the fidelities. In Figs.~\ref{F9}(b) and \ref{F9}(d), we plot the fidelities as the functions with respect to $t_f$ and $\gamma$ for the state transfer and the entanglement generation, respectively. Also clearly, when the atomic spontaneous emissions exist, the influence of the atomic spontaneous emissions on the fidelities can be restrained by a longer $t_f$. Nevertheless, for the experimental feasibility and the efficiency of the scheme, we pick an appropriate pair of values $g=10\Omega_0$ and $t_f=40/\Omega_0$.
\begin{figure}[htb]
\begin{center}
\includegraphics[width=\linewidth]{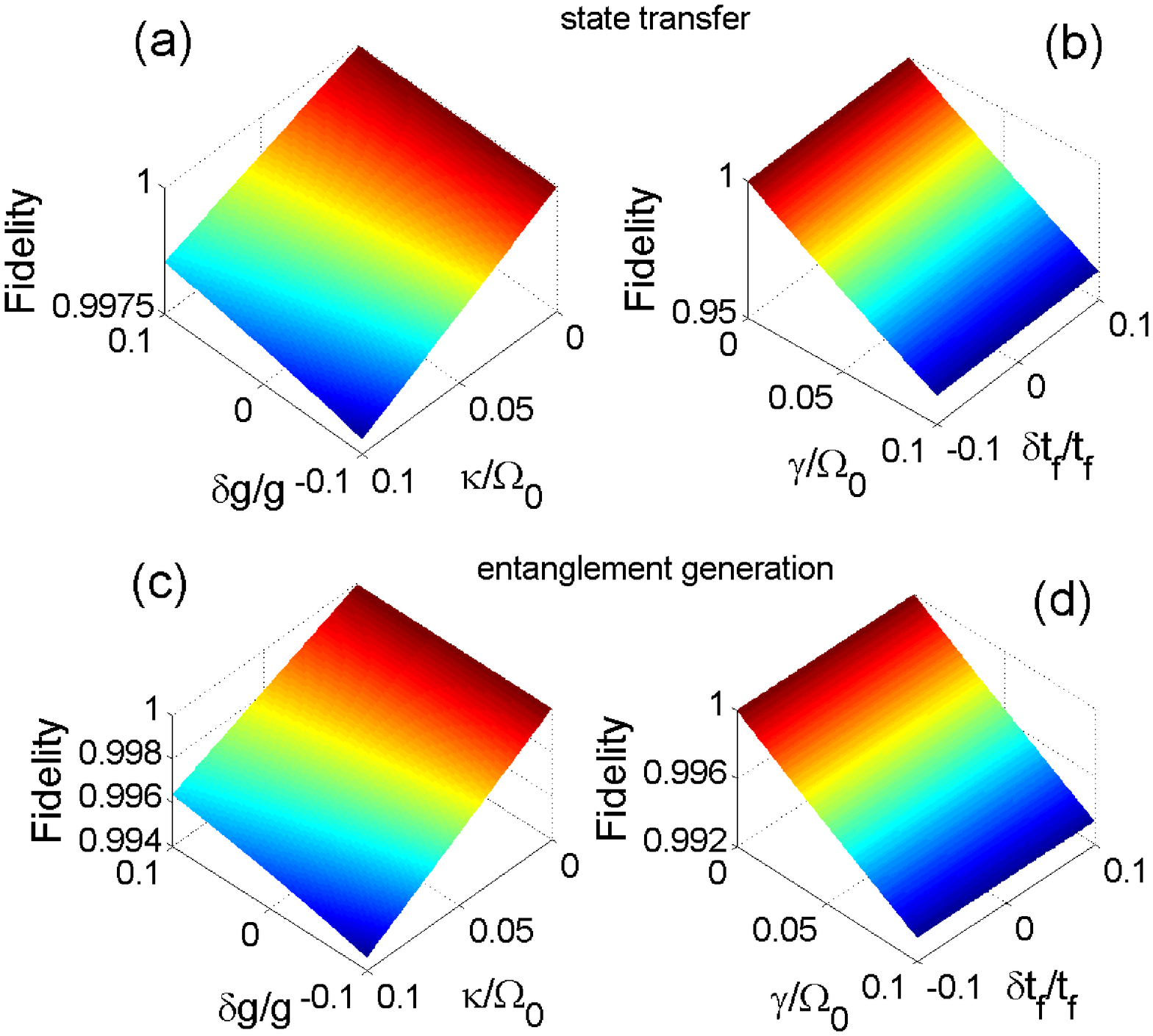}\\
\caption{(Color online) (a) and (c): the fidelities versus $\kappa/\Omega_{0}$ and the variations in $g$ with $\gamma=0$; (b) and (d): the fidelities versus $\gamma/\Omega_{0}$ and the variations in $t_f$ with $\kappa=0$. The parameters used here are the same as in Fig.~\ref{F6}.}\label{F10}
\end{center}
\end{figure}

Experimentally, it is too hard to have the theoretically predicted values of the parameters. Therefore, in Fig.~\ref{F10}, we give the numerical simulations to discuss the joint effects of the photon leakages from the cavity and the variations in the atom-cavity coupling strength $g=10\Omega_{0}$ and those of the atomic spontaneous emissions and the variations in the operation time $t_f=40/\Omega_{0}$ on the fidelities. Figures~\ref{F10}(a) and \ref{F10}(c) also indicate that when the photon leakages from the cavity exist, a bigger value of $g$ can greatly depress the influence of the photon leakages from the cavity on the fidelities. Although either $\kappa$ or the variations in $g$ can not be controlled in experiment, the fidelities are over 0.99 even under the terrible condition \{$\delta g/g=-0.1$, $\kappa=0.1\Omega_0$\}. Figures~\ref{F10}(b) and \ref{F10}(d) show that the fidelities are almost not affected by the variations in $t_f=40/\Omega_{0}$ whenever the atomic spontaneous emissions exist or not. In other words, the dressed-state scheme are extremely robust against the variations in the chosen operation time. In addition, once the operation time is determined, the effects of the atomic spontaneous emissions on the fidelities are independent on the variations in the operation time.

\section{Conclusion}\label{c}
In conclusion, we have developed the dressed-state method to achieve the fast adiabatic quantum state transfer and entanglement generation between two $\Lambda$-type atoms trapped in an optical cavity. There is no a direct coupling of the target state and the initial state appearing in the Hamiltonian. The pulses are modified with the high experimental feasibility and can be smoothly turned on or off, which ensure the feasibility of the scheme in practice. During the whole evolution, the adiabatic condition is not necessary to be met, and thus even within a very short operation time ,the state transfer and the entanglement generation can be achieved with quite high fidelities. The introductions of the Zeno limit condition and the auxiliary function $G(t)$ restrain the populations of all of the excited states and hence the scheme is robust against the decoherence induced by the atomic spontaneous emissions and the photon leakage from the cavity. Besides, the results of the numerical simulations show that the dressed-state scheme is robust against the errors of the generated pulse sequences and the variations in the chosen parameters.

Based on Ref.~\cite{STK2005}, by using cesium atoms and a set of cavity QED predicted parameters $(g,\kappa,\gamma)/2\pi=(750,3.3,2.62)$ MHz, we can achieve the two-atom quantum state transfer and maximum entanglement generation with the fidelities $F=0.985$ and $F=0.996$, respectively. Therefore, it allows us to construct an atomic system for the quantum state transfer and the entanglement generation in the presence of decoherence. In a word, by using dressed states, we have implemented the fast, feasible and robust two-atom adiabatic quantum state transfer and entanglement generation. In the further work, it could be interesting to apply the dressed-state method to more complex systems for preparing more complex entanglement and constructing quantum gates.
\\
\begin{center}
{\bf{ACKNOWLEDGMENT}}
\end{center}

The authors would like to express their sincere gratitude and thanks to the unanimous referee for her/his positive and critical comments which helped in improving the presentation of the work. The authors are grateful to X.Q. Shao, S.L. Su, and Q.C. Wu for useful discussions. This work was supported by the National Natural Science Foundation of China under Grants No. 11464046 and No. 61465013.\\
\begin{center}
{\bf{APPENDIX: TIME-INDEPENDENT ADIABATIC FRAME HAMILTONIAN AND DRESSED-STATE FRAME HAMILTONIAN}}
\end{center}

The contents in this appendix have been briefly explained in the work of Baksic \emph{et~al.}~\cite{AHA2016}, and here we give a more detailed description.

The pulses in the Hamiltonian~(\ref{e6}) can be parametrized by the frequency $\Omega(t)$ and the angle $\theta(t)$
\begin{equation*}\label{eA1}
\Omega_1(t)=-\Omega(t)\sin\theta(t),\quad\Omega_2(t)=\Omega(t)\cos\theta(t), \tag*{(A1)}
\end{equation*}
with $\Omega(t)=\sqrt{\Omega_1(t)^2+\Omega_2(t)^2}$ and $\theta(t)=\arctan(\Omega_1(t)/\Omega_2(t))$, and we can easily obtain the time-dependent eigenstates of $H(t)$
\begin{align*}\label{eA2}
|\varphi_d(t)\rangle&=\cos\theta(t)|\phi_1\rangle+\sin\theta(t)|\phi_5\rangle,\\
|\varphi_{\pm}(t)\rangle&=\frac{1}{\sqrt{2}}\Big(\sin\theta(t)|\phi_1\rangle\mp|\phi_d\rangle-\cos\theta(t)|\phi_5\rangle\Big),\tag*{(A2)}
\end{align*}
with the eigenvalues $E_d=0$ and $E_{\pm}=\pm \Omega(t)$, respectively.

Then move the time-dependent adiabatic frame to the time-independent adiabatic frame by the unitary operator $U(t)=\sum_{j=d,\pm}|\varphi_j\rangle\langle\varphi_j(t)|$. In the time-independent adiabatic frame the Hamiltonian~(\ref{e6}) becomes
\begin{equation*}\label{eA3}
H_{\rm ad}(t)=\Omega(t)M_z+\dot{\theta}(t)M_y, \tag*{(A3)}
\end{equation*}
where $M_z=|\varphi_+\rangle\langle\varphi_+|-|\varphi_-\rangle\langle\varphi_-|$ and $M_y=i(|\varphi_+\rangle+|\varphi_-\rangle)\langle\varphi_d|/\sqrt{2}+\rm H.c.$. The second term of the Hamiltonian~\ref{eA3} corresponds to the nonadiabatic couplings which may lead to an imperfect state transfer.

Moving the modified Hamiltonian
\begin{align*}\label{eA4}
H_{\rm mod}(t)&=H(t)+H_c(t)\nonumber\\
&=\Omega'_1(t)|\phi_1\rangle\langle\phi_d|+\Omega'_2(t)|\phi_5\rangle\langle\phi_d|+\rm H.c., \tag*{(A4)}
\end{align*}
to the frame defined by $V(t)$, we obtain the new dressed-state frame Hamiltonian
\begin{align*}\label{eA5}
H_{\rm new}(t)&=V H_{\rm ad}(t)V^\dag+VU H_c(t)U^\dag V^\dag+i\frac{d V}{dt}V^\dag\nonumber\\
&=\eta(t)\Big(|\tilde{\varphi}_+(t)\rangle\langle\tilde{\varphi}_{+}(t)|-|\tilde{\varphi}_{-}(t)\rangle\langle\tilde{\varphi}_{-}(t)|\Big)\nonumber\\
&\quad+\xi(t)\Big[\Big(|\tilde{\varphi}_+(t)\rangle-|\tilde{\varphi}_{-}(t)\rangle\Big)\langle\tilde{\varphi}_{d}(t)|+\rm H.c.\Big], \tag*{(A5)}
\end{align*}
with the time-dependent parameters
\begin{align*}\label{eA6}
\eta(t)&=[g_z(t)+\Omega(t)]\cos\mu(t)-\dot{\theta}(t)\sin\mu(t)\nonumber\\
\xi(t)&=\frac{1}{\sqrt{2}}\Big\{i[g_z(t)+\Omega(t)]\sin\mu(t)+i\dot{\theta}(t)\cos\mu(t)+[\dot{\mu}(t)-g_x(t)]\Big\}. \tag*{(A6)}
\end{align*}
When the control parameters are chosen as
\begin{equation*}\label{eA7}
g_x(t)=\dot{\mu}(t),\quad g_z(t)=-\Omega(t)-\frac{\dot{\theta}(t)}{\tan\mu(t)}, \tag*{(A7)}
\end{equation*}
the second term of the Hamiltonian~\ref{eA5} is removed. In other words, the correction Hamiltonian $H_c(t)$ has been designed such that it cancels the unwanted off-diagonal elements in $H_{\rm new}(t)$.

\end{document}